\documentclass[preprint,amsmath,amssymb,aps,prb,groupedaddress]{revtex4-1}

\usepackage{graphicx}
\usepackage{dcolumn}
\usepackage{bm}
\usepackage[colorlinks, linkcolor={blue}, urlcolor={blue}, citecolor={blue}]{hyperref}

\begin{document}



\title{Search for pressure-induced tricriticality in Cr}

\author{A. Schade}
\email[Email: ]{a.schade@tum.de}
\homepage[ORCID: ]{http://orcid.org/0000-0002-6584-770X}
\affiliation{Physik-Department E21, Technische Universit{\"a}t M{\"u}nchen, 85748 Garching, Germany}
\author{T. Adams}
\affiliation{Physik-Department E21, Technische Universit{\"a}t M{\"u}nchen, 85748 Garching, Germany}
\author{A. Chac\'on}
\affiliation{Physik-Department E21, Technische Universit{\"a}t M{\"u}nchen, 85748 Garching, Germany}
\author{R. Georgii}
\affiliation{Physik-Department E21, Technische Universit{\"a}t M{\"u}nchen, 85748 Garching, Germany}
\affiliation{Heinz Maier-Leibnitz Zentrum (MLZ), Technische Universit{\"a}t M{\"u}nchen, 85748 Garching, Germany}
\author{C. Pfleiderer}
\affiliation{Physik-Department E21, Technische Universit{\"a}t M{\"u}nchen, 85748 Garching, Germany}
\author{P. B{\"o}ni}
\affiliation{Physik-Department E21, Technische Universit{\"a}t M{\"u}nchen, 85748 Garching, Germany}

\date{\today}

\begin{abstract}
The antiferromagnetic ordering of chromium has long been known for its peculiar physical properties. One of them is the observation of the weak first-order character of the N\'eel transition that is explained by the lack of a stable fixed point by Bak and Mukamel.
Barak et al. predicted that by lowering the symmetry of the order parameter by the application of uniaxial pressure along the $[110]$ direction changes the N{\'e}el transition to second-order. In a previous experiment by Fawcett et al., irreversible broadening of the N{\'e}el transition was already observed between $160\,\mathrm{bar}$ and $300\,\mathrm{bar}$, which could, however, be caused by plastic deformation. Using an improved setup with reduced stress inhomogeneities we succeeded to increase the pressure range until irreversible broadening is observed above $450\,\mathrm{bar}$. Despite the observed tripling of the intensity of the magnetic Bragg peak $[0, 0, 1 - \delta]$ at $p_{[110]} \ge 450\,\mathrm{bar}$, indicating a single ${\bf Q}_\pm$-domain state of the sample, no hints for a tricritical point were observed.
\end{abstract}


\maketitle




\section{\label{sec:level1Intro}Introduction}

Chromium exhibits a fascinating magnetic phase diagram despite its simple body-centered cubic structure. Nesting of the Fermi surface leads to the formation of a sinusoidal incommensurate spin-density wave with ordering vectors ${\bf Q}_\pm = (2\pi/a)(0,0,1\pm\delta)$, where $a = 2.88$ \AA\ is the lattice constant and $\delta = 0.048$ at $T = 100\, {\rm K}$. \cite{BIBFawcettSummary} A spin-flop transition from the longitudinal (LSDW) to the transverse spin-density wave (TSDW) phase takes place at $T_{SF} = 121$ K. The incommensurate ordering leads to a complicated spectrum of magnetic excitations which emerges from the magnetic satellites and is still not understood\cite{PeterInelasticNeutronA,PeterInelasticNeutronB} despite intense experimental and theoretical efforts.\cite{fishman1996} 

\begin{figure}[htb]
\centering
\includegraphics{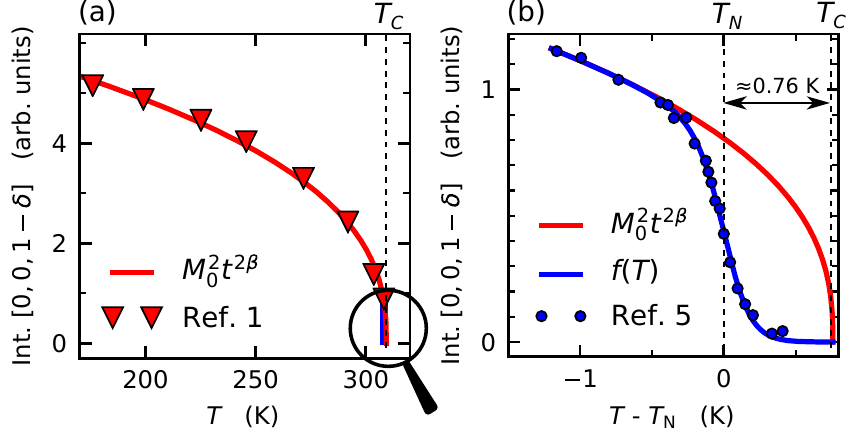}
\caption{\label{NEUTRONfigNeelTwoScales} $T$-dependence of the $(0,0,1-\delta)$ magnetic satellite peak $I(T)$ of Cr. \cite{BIBFawcettSummary} (a) On a coarse $T$-scale, the data points can be parametrized assuming a second-order phase transition with $I(T) \propto t^{2\beta}$ (red line). (b) The enlargement of $I(T)$ close to $T_N$ reveals a weak
first-order phase transition. \cite{BIBFawcettUniax} The blue solid line is explained in paragraph \ref{sec:level1ExpResults}.}
\end{figure}

In comparision to conventional ferro- and antiferromagnets, which enter the paramagnetic phase via a continuous second-order phase transition, Cr exhibits a weak first-order phase transition similar to the helimagnetic to paramagnetic transition in MnSi. Fig. \ref{NEUTRONfigNeelTwoScales}~(a) depicts the temperature dependence of the intensity of the magnetic satellite peak near the N\'eel temperature $T_N$ the intensity of which is proportional to the square of the staggered magnetisation $M$. \cite{BIBFawcettSummary}
Assuming a second-order phase transition, the data points are fitted with the expression $M^2 = M_0^2 t^{2 \beta}$, where the reduced temperature is given by $t = (T_c -T)/T_c$, yielding $\beta = 0.19\pm0.01$ and a putative transition temperature $T_c = T_N + (0.76\pm0.14)\, {\rm K}$, which exceeds the experimentally determined $T_N$. Indeed, zooming into the region near $T_N$ reveals that the phase transition is weakly first-order as indicated by the blue data points shown in panel (b) of Fig. \ref{NEUTRONfigNeelTwoScales}. \cite{BIBFawcettUniax}

Bak and Mukamel argued that fluctuations modify the order of the transition \cite{BakMukamelRenormalization} similarly as observed in MnSi, where the first-order phase transition from the helimagnetic phase to the paramagnetic phase is explained in terms of a Brasovskii-transition\cite{janoschek2013} which is induced by the large phase space available for the fluctuations and concomitant interactions.\cite{brazovskii1975} Based on the Landau-Ginsburg-Wilson Hamiltonian proposed by Bak and Mukamel, Barak and Walker predicted that applying uniaxial pressure along a $[110]$ direction in Cr the symmetry of the order parameter is lowered, and thus the available phase space for fluctuations is reduced. Therefore, the order of the N\'eel transition would change from first- to second-order and a tricritical point is expected at a critical pressure $p_{[110]} = p_{[110]}^{\mathrm{crit}}$. \cite{BIBZviBarak}


A first attempt to investigate the N{\'e}el-transition of Cr under uniaxial pressure was made by Fawcett et al. \cite{BIBFawcettUniax} However, the prediction of Barak et al. could neither be verified nor falsified because the N\'eel transition began to broaden irreversibly for pressures in the range $160\,\mathrm{bar} \le p_{[110]} \le 300\,\mathrm{bar}$. Fawcett et al. conjectured that the broadening was caused by the formation of residual stresses due to the occurrence of plastic deformation.\cite{BIBFawcettUniax}

In order to reduce the plastic deformation under the application of uniaxial pressure, we have developed an improved experimental setup and repeated the experiment of Fawcett et al. The uniaxial pressure cell was developed by Chac{\'o}n et al.\cite{AlfonsoBachelorThesis} based on work reported in Refs. \onlinecite{pfleiderer1997,waffenschmidt1999}.
It was successfully applied to characterize the dependence of the magnetic order of MnSi on uniaxial pressure.\cite{ChaconUniaxMnSi} Adams et al. used the same pressure cell in an experiment on Cr, where the uniaxial pressure was applied along the $[001]$ crystallographic axis.\cite{BIBtimChromium}

The value of the critical pressure $p_{[110]}^{\mathrm{crit.}}$ is not predicted by Barak's model due to unknown material constants. However, we believe that applying a uniaxial pressure of the order of $0.6\,\mathrm{kbar}$ along the $[110]$ direction is sufficient to suppress the $4$ magnetic Bragg peaks $[\pm \delta, \pm \delta, 1]$, because it has been shown by Adams et al. that the application of $0.6\,\mathrm{kbar}$ along a $[001]$ direction achieves complete suppression of the SDW propagation direction along $[001]$.\cite{BIBtimChromium}
Because the suppression of magnetic Bragg peaks is supposed to coincide with the extinction of the associated fluctuations, we conjecture that $p_{[110]}^{\mathrm{crit.}} \approx 0.6\,\mathrm{kbar}$.


\section{\label{sec:level1ConseqStress}Implications of stress inhomogeneities}

Uniaxial pressure cells are a formidable tool to study effects of magnetoelastic coupling in antiferromagnetic materials.\cite{haelgEtAlNew,BIBFawcettUniax} As discussed above, the application of $0.6\,\mathrm{kbar}$ along the $[001]$ direction may be sufficient to completely suppress the magnetic Bragg peaks $[0, 1, \pm \delta]$.\cite{BIBtimChromium} This pressure corresponds to the application of only $60\,\mathrm{N}$ on a cross sectional area of $1\,\mathrm{mm}^2$. In contrast, even a very large external magnetic field of $7\,\mathrm{T}$ fails to affect the order parameter in Cr due to the very weak coupling between the magnetic field and the spin density wave.\cite{BIBFawcettUniax}

A disadvantage of applying uniaxial pressure in comparison to an externally applied magnetic field is the presence of pressure inhomogeneities in the sample, which are difficult to control. A discussion of the dependence of the stress inhomogeneities as a function of the dimensions of the sample and the mechanical deficiencies of the pressure cells may be found in Ref. \onlinecite{MSc-Schade}.

Typically, pressure inhomogeneities lead to a smearing of phase transitions. Using the experimentally determined value $dT_N/d\sigma = - 1.5\pm0.4$ K/kbar for Cr, \cite{BIBFawcettUniax} the shift of the transition temperature $T_c - T_N = 0.76\, {\rm K}$ corresponds to a pressure $p_{[110]} = 510\, {\rm bar}$. Therefore, it has to be made sure that the standard deviation $\sigma_p$ of the uniaxial pressure within the sample is $\sigma_p \ll 510\, {\rm bar}$.

Plastic deformation of a sample sets in when the von Mieses stress, $\sigma_{{mis}} := [\frac{1}{2}((\sigma_{{xx}} - \sigma_{{yy}})^2 + (\sigma_{{yy}} - \sigma_{{zz}})^2 + (\sigma_{{zz}} - \sigma_{{xx}})^2) + 3 (\sigma_{{xy}}^2 + \sigma_{{xz}}^2 + \sigma_{{yz}}^2)]^{1/2}$, exceeds the yield strength, $\sigma_{{yield}}$.\cite{MathTheoPlastic} For typical sample dimensions, all components except $\sigma_{{zz}}$ may be neglected \cite{mySecondPaper} yielding $\sigma_{{mis}} \approx |p_{[110]}|$. However, stress inhomogeneities lower the threshold for plastic deformation. On a $2\sigma_p$ tolerance level, plastic deformation starts when the mean pressure $\bar p_{[110]}$ exceeds the threshold $\sigma_{{yield}} - 2 \sigma_p$.


\section{\label{sec:level1ExpSetup}Experimental Setup}

\begin{figure}[htb]
\centering
\includegraphics{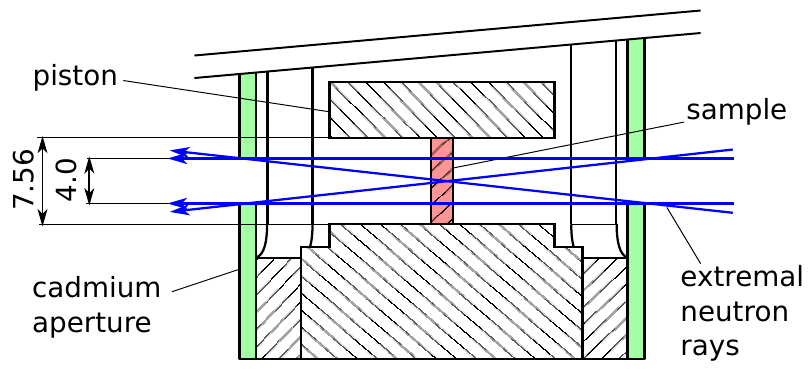}
\caption{Cross-sectional view of the lower part of the pressure cell. The sample is indicated by the vertical rectangle (light red). Cadmium sheets (light green) have been wrapped around the sample cage, and leave windows with a height of $4\,\mathrm{mm}$. The window clips the St. Venant region.\cite{mySecondPaper} %
}
\label{NEUTRONfigCadmiumAperture}
\end{figure}

The Cr sample we studied had the shape of a prism of height $h = 7.56\,\mathrm{mm}$ and a square cross section with sides $a = 2\,\mathrm{mm}$.
One component of the stress errors is caused by barreling. It can be shown that these errors concentrate near the contact surface, and become exponentially weaker as a function of the $z$-coordinate. \cite{mySecondPaper} Our choice of $h/l$ guarantees that average stress errors caused by barrelling are below $10\%$, if boundary regions with a height of at least $1.6\,\mathrm{mm}$ are clipped using cadmium apertures. Exceeding this estimate we used apertures that reduce the neutron beam by approximately $1.8\,\mathrm{mm}$ on each side (Fig. \ref{NEUTRONfigCadmiumAperture}).

Another important component of the stress errors is caused by bending of the sample, which is caused by non-parallel contact surfaces if standard, guided pistons are used. In order to minimize stress inhomogeneities caused by non-parallel contact surfaces we used a piston that compensates inclinations as proposed by Schade and B\"oni using grease as lubricant. \cite{mySecondPaper} For our design of the piston that compensates for inclinations we postulate that the bending of the sample mostly depends on the centering of the sample in the apparatus, while surface inclination is of lesser importance. We estimate that the centering error of the sample is bounded by $\Delta d < 0.2\,\mathrm{mm}$, and thus the stress error caused by eccentricity is below $2\%$.
The above considerations suggest a total error in the stress of $\approx 12\%$.
Nevertheless, our setup cannot guarantee that the stress errors are less than $12\%$, since other factors could contribute.\cite{mySecondPaper}

\begin{figure}[htb]
\centering
\includegraphics[width=70mm]{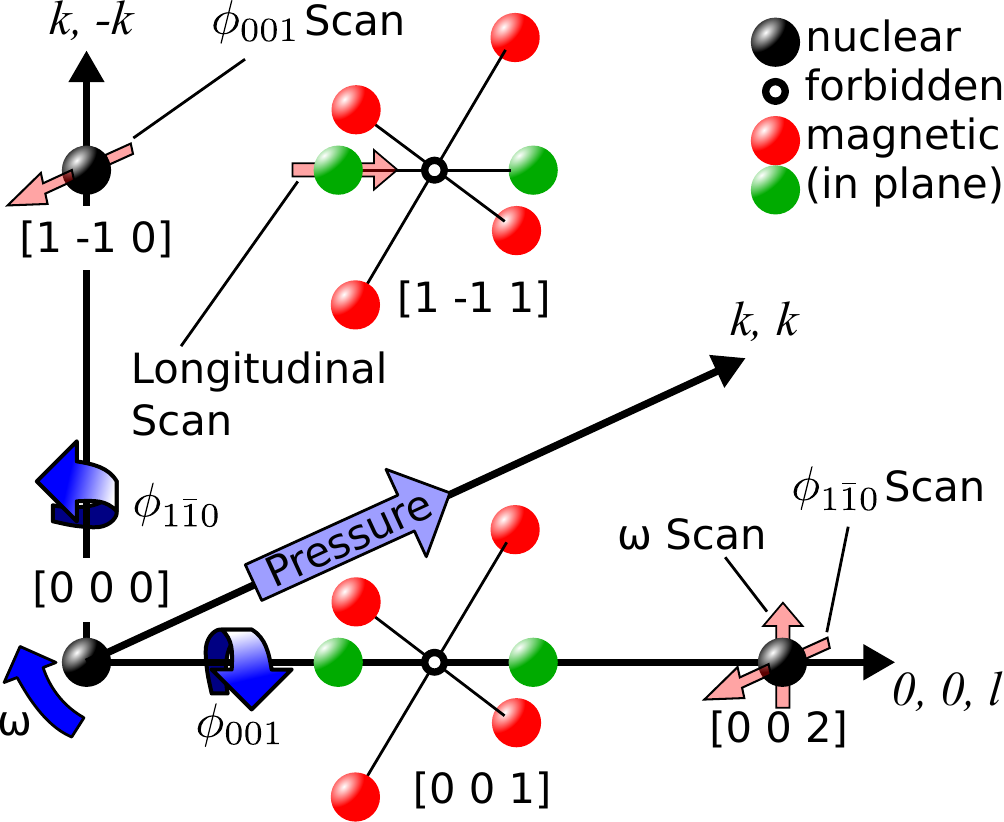}
\caption{Scattering plane in reciprocal space. The uniaxial pressure is applied along $[110]$, i.e. normal to the scattering plane. %
The longitudinal scan is used to measure the magnetization as a function of temperature, and the $\omega$, $\phi_{1 \bar{1} 0}$, and $\phi_{0 0 1}$ scans are used for realigning the sample after each pressure change. %
}
\label{NEUTRONfigPerformedScans}
\end{figure}

Elastic neutron scattering experiments were conducted at MIRA (FRM II)
\cite{georgii2018,MIRAtechnicalPaper}
in the triple axis mode with an energy transfer $\Delta E = 0$. Uniaxial pressure was applied along the $[110]$ direction using the pressure cell of Chac{\'o}n.\cite{AlfonsoBachelorThesis} We performed $12$ temperature sweeps during which we repeatedly measured the integral intensity of the magnetic Bragg peak $[0, 0, 1 - \delta]$ using a longitudinal scan in reciprocal space as shown in Fig. \ref{NEUTRONfigPerformedScans}. Each sweep started at a temperature $T_1 > T_{N}$ followed by a decrease of $T$ at a constant rate of $2.7$ K/h to $T_2 < T_{N}$.

After each pressure change, we optimized the orientation of the sample to compensate a possible tilting of the crystal planes due to a rigid rotation or a plastic deformation by adjusting the angles $\omega$, $\phi_{1 \bar{1} 0}$, and $\phi_{0 0 1}$ (see Fig. \ref{NEUTRONfigPerformedScans}). For this purpose the intensity of the nuclear Bragg peaks $[1, -1, 0]$ and $[0, 0, 2]$ was optimized. 

\begin{figure}[htb]
\centering
\includegraphics[trim=0 3mm 0 0]{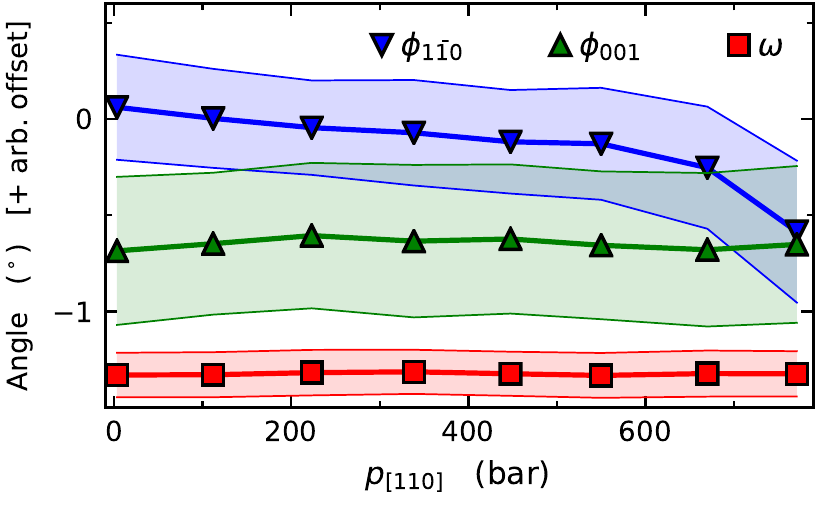}
\caption{\label{fig:fwhm}Orientation of the scattering plane parameterized by the angles $\phi_{1 \bar{1} 0}$, $\phi_{0 0 1}$, and $\omega$ as a function of the applied uniaxial pressure. The error bars are smaller than the symbol size. The shading indicates the full width at half maximum (FWHM) of the Bragg peaks used for the alignment (Fig.~\ref{NEUTRONfigPerformedScans}).}
\end{figure}

Fig. \ref{fig:fwhm} shows the optimized goniometer angles $\omega$, $\phi_{1 \bar{1} 0}$, and $\phi_{0 0 1}$ for each pressure. It is seen that the angles $\omega$ and $\phi_{0 0 1}$ do not change significantly. In contrast, $\phi_{1 \bar{1} 0}$ decreases strongly by about $0.5^\circ$ for $p_{[110]} \ge 550\, {\rm bar}$. We will see below that the strong tilting of the $(001)$ crystal planes about the $[{1 \bar{1} 0}]$ axis is linked to the observed plastic deformation of the crystal.


\section{\label{sec:level1ExpResults}Experimental Results}

\begin{figure}[htb]
\centering
\includegraphics{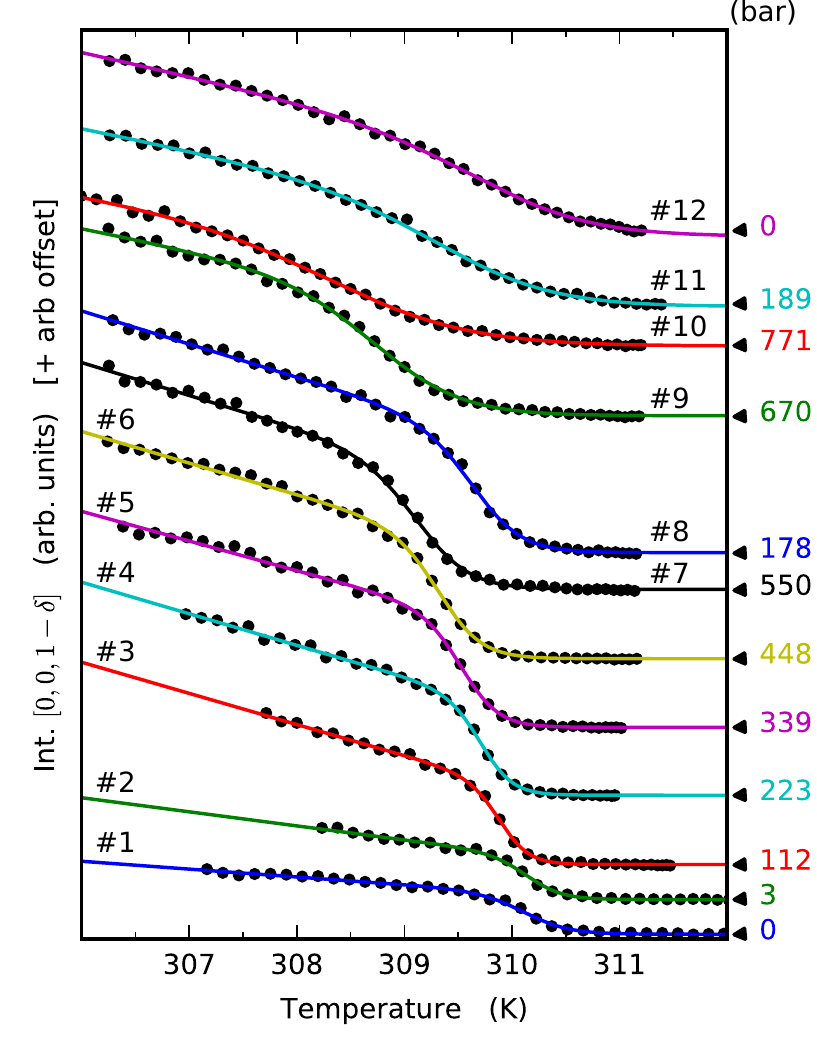}
\caption{Summary of all temperature sweeps performed. The numbering \#1 to \#12 reflects the sequence of the measurements. The data sets are shifted vertically by a constant for better visibility. The scaling of the intensity axis is the same for each curve. The error bars are smaller than the size of the symbols. The solid lines depict the fitted function $f(T)$ given by Eq.~(\ref{NEUTRONrealFitFunc}). The numbers of the axis on the right hand side indicate the applied pressure in bar.}
\label{NEUTRONfigAllFits}
\end{figure}

Fig. \ref{NEUTRONfigAllFits} summarizes the 12 temperature sweeps we performed. They are numbered in the sequence of occurence. From \#1 to \#7 the uniaxial pressure was monotonically increased from $0\, {\rm bar}$ to $550\, {\rm bar}$. In run \#1 the piston had no mechanical contact with the sample. The data shows that the smearing of the phase transition does not change significantly. However, the step height increases by approximately a factor of 3 indicating a complete population of the SDW-domain with $\bf Q_\pm$ parallel to $[001]$. There is also an indication for a lowering of $T_N$ by $\approx 0.8\, {\rm K}$ with increasing pressure quantitatively consistent with the known pressure dependence of $T_N$. 

Subsequently decreasing the pressure from $550\, {\rm bar}$ to $178\, {\rm bar}$ (run \#8) essentially resulted in a shift of the data of run \#7 to higher temperature without involving a significant change of the profile. This result indicates that (i) the sample behaved reversibly, and (ii) that the sample remained in a single $\bf Q_\pm$ state. During the course of runs \#9 and \#10 the pressure was increased further before it was reduced again to $0\, {\rm bar}$ (\#11, \#12). The data clearly show a significant broadening of the transition that is irreversible.  We conclude that  the sample was plastically deformed.

\begin{figure}[htb]
\centering
\includegraphics[trim=0 3mm 0 0]{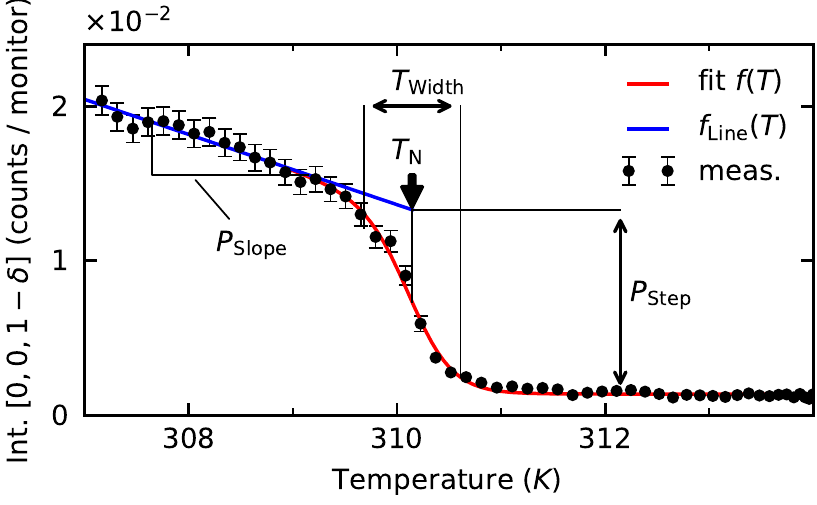}
\caption{\label{fig:illustrfit}Fit of the data set \#1 (0 bar) by $f(T)$ from Eq. (\ref{NEUTRONrealFitFunc}). The width of the transition, $T_{\mathrm{Width}}$, is approximately given by the $[10\% \cdot P_{\mathrm{Step}}, 90\% \cdot P_{\mathrm{Step}}]$ levels. The error bars represent the confidence interval of $2 \sigma$. They are not shown if they are smaller than the size of the symbols.
}
\end{figure}

For a quantitative estimate we parametrized the data by the function $f(T)$ given by
\begin{align}
	\label{NEUTRONrealFitFunc} f(T) &= f_{\mathrm{Line}}(T) \cdot f_{\mathrm{Sigmoid}}(T) + P_{bg}
\end{align}
where
\begin{align}
	f_{\mathrm{Line}}(T) &:= P_{\mathrm{Slope}} \cdot (T - T_{N}) + P_{\mathrm{Step}} \\
	f_{\mathrm{Sigmoid}}(T) &:= \frac{1}{1 + \exp\left( \frac{T - T_{N}}{\Delta T} \right)} \\
	\Delta T &:= \frac{T_{\mathrm{Width}}}{2 \log 9}.
\end{align}
Here, $P_{bg}$, $P_{\mathrm{Slope}}$, $P_{\mathrm{Step}}$, $T_{N}$ and $T_{\mathrm{Width}}$ are the fit parameters designating the background, the linear slope of the integrated intensity of the magnetic peak, the increase of the intensity at the phase transition, the N\'eel temperature, and the width of the transition, respectively (see Fig. \ref{fig:illustrfit}). The choice of $f_{\mathrm{Sigmoid}}(T)$ in the form of a Fermi function is purely empirical.

\section{\label{sec:level1Discuss}Discussion}

\begin{figure}[htb]
\centering
\includegraphics[trim=0 3mm 0 0]{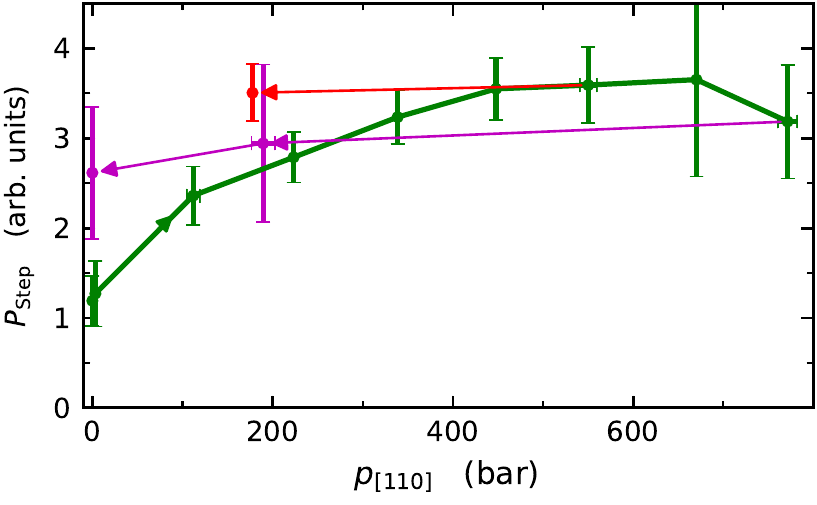}
\caption{Increase of the intensity of the magnetic satellite peak $(0,0,1-\delta)$ at $T_N$ versus $p_{[110]}$. With increasing $p_{[110]}$, $P_{\mathrm{Step}}$ increases. After reaching $550\, {\rm bar}$ and $771\, {\rm bar}$, $p_{[110]}$ was reduced to $178\, {\rm bar}$ (red arrow) and $0\, {\rm bar}$ (violett arrow), respectively. Irreversibilities are observed for $p_{[110]} \ge 550\, {\rm bar}$.
}
\label{NEUTRONfigPStep}
\end{figure}

The pressure dependence of $P_{\mathrm{Step}}$ shown in Fig. \ref{NEUTRONfigPStep} confirms the conclusions drawn from the raw data notably that a single $\bf Q_\pm$-state is populated at a high pressure of $p_{[110]} \approx 448\, {\rm bar}$. There is an indication of a downturn of $P_{\mathrm{Step}}$ at $771\, {\rm bar}$ which does not recover when the pressure is released suggestive of plastic deformation.

\begin{figure}[htb]
\centering
\includegraphics[trim=0 3mm 0 0]{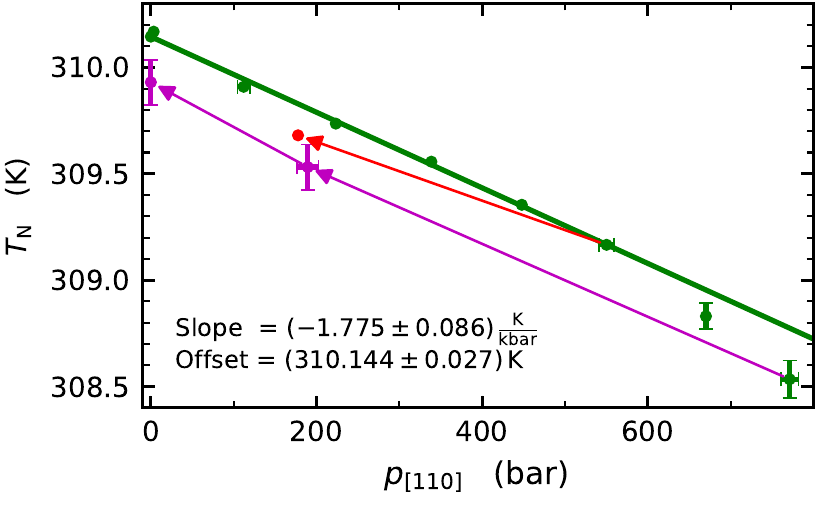}
\caption{N{\'e}el temperature $T_{N}$ versus $p_{[110]}$. After reaching $550\, {\rm bar}$ and $771\, {\rm bar}$, $p_{[110]}$ was reduced to 178 bar (red arrow) and $0\, {\rm bar}$ (violett arrow), respectively. Irreversibilities are observed for $p_{[110]} \ge 550\, {\rm bar}$. Error bars are omitted when they are smaller than the symbol size.}
\label{NEUTRONfigTNeel}
\end{figure}

Fig. \ref{NEUTRONfigTNeel} shows that $T_N$ decreases linearly as long as $p_{[110]} \le 550\,\mathrm{bar}$.  Above $550\, {\rm bar}$, $T_N$ decreases faster indicating the development of stress inhomogeneities, i.e. $\sigma_p > 0$. A linear regression of the slope yields
\begin{align}
	\label{FinalResultDTNeelQuant} &
	\frac{dT_{N}}{dp} = (-1.77 \pm 0.10)\,\frac{\mathrm{K}}{\mathrm{kbar}}.
\end{align}
In addition to the statistical error of the fit, a systemic error of $0.012\,\mathrm{K}/\mathrm{kbar}$ due to the pressure calibration is included.

The slope as determined in our experiment is in good agreement with the slope determined by Fawcett et al. who obtained $dT_N/dp = -1.5\pm0.4\, {\rm K/kbar}$.\cite{BIBFawcettUniax} It compares also well with the slope as reported by McWhan and Rice\cite{McWhanPressure} under application of hydrostatic pressure who found $(dT_{N}/dp)_{iso} = -5.1\,\frac{\mathrm{K}}{\mathrm{kbar}}$ which is $\simeq 2.9$ times larger than our value, i.e. very close to the theoretical value of 3.

\begin{figure}[htb]
\centering
\includegraphics[trim=0 3mm 0 0]{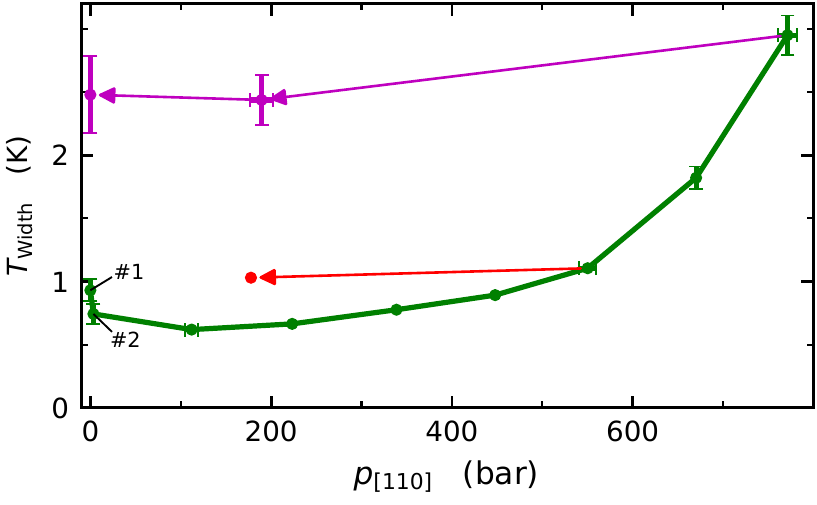}
\caption{\label{NEUTRONfigTWidth}Width of the N{\'e}el transition, $T_{\mathrm{Width}}$, versus $p_{[110]}$. After reaching $550\, {\rm bar}$ and $771\, {\rm bar}$, $p_{[110]}$ was reduced to $178\, {\rm bar}$ (red arrow) and $0\, {\rm bar}$ (violett arrow), respectively. Irreversibilities are observed for $p_{[110]} \ge 550\, {\rm bar}$. Error bars are omitted when they are smaller than the symbol size. The data points with the labels \#1 and \#2 correspond to the sweeps \#1 ($0\, {\rm bar}$) and \#2 ($3\, {\rm bar}$).}
\end{figure}

Finally, Fig. \ref{NEUTRONfigTWidth} shows the central result of our investigation, namely the pressure dependence of the width of the transition. Initially, $T_{\mathrm{Width}}$ shrinks significantly from $0.93\,\mathrm{K}$ at $0\, {\rm bar}$ (sweep \#1) to $0.62\,\mathrm{K}$ at $3\, {\rm bar}$ before increasing again and assuming at $p_{[110]} \approx 448\,\mathrm{bar}$ the value of sweep \#1. At larger pressures $p_{[110]} \ge 550\, {\rm bar}$, $T_{\mathrm{Width}}$ increases rapidly. When releasing $p_{[110]}$ in this regime, $T_{\mathrm{Width}}$ maintains its high value suggesting plastic deformation. The signature of a tricritical point, as predicted by Barak et al.\cite{BIBZviBarak}, would be reflected in a marked and {\sl reversible} increase of $T_{\mathrm{Width}}$ as typically observed for a continuous phase transition. This is not observed.

In the experiment reported by Fawcett et al.\cite{BIBFawcettUniax}, the width of the transition was found to increase by approximately a factor of two between $160\, {\rm bar}$ and $300\, {\rm bar}$, in contrast to our experiment, where a significant broadening occurs only above $550\, {\rm bar}$. We attribute the behavior of our sample to a significantly more homogeneous stress distribution.

Finally, let us discuss the width of the magnetic phase transition in Cr at zero pressure. The data points labeled with \#1 and \#2 in Fig. \ref{NEUTRONfigTWidth} correspond to sweeps \#1 ($p_{[110]} = 0\, {\rm bar}$) and \#2 ($p_{[110]} = 3\, {\rm bar}$). As soon as the piston comes into contact with the sample, $T_{\mathrm{Width}}$ is significantly reduced by about $0.2\,\mathrm{K}$ while a significant difference of $T_{\mathrm{N}}$ and $T_{\mathrm{Step}}$ between sweeps \#1 and \#2 is not observed. The decrease of $T_{\mathrm{Width}}$ may be a consequence of an improved alignment of $\bf Q_\pm$ perpendicular to $p_{[110]}$ due to the applied pressure.
According to Eq. (\ref{FinalResultDTNeelQuant}) a pressure difference $\triangle p \simeq 110\, {\rm bar}$ is required to shift $T_N$ by $0.2\, {\rm K}$ which is much larger than the applied pressure of $3\, {\rm bar}$ in sweep \#2. Therefore, the width of the transition at $T_N$ at zero pressure, $T_{\mathrm{Width}}\approx 0.93\,{\rm K}$, is intrinsic (maybe caused by imperfections of the crystal) and is not caused by residual stress in the sample.


\section{\label{sec:level1Concl} Summary and Outlook}


Based on our experimental results, namely that the transition at $T_N$ remains 1$^{\rm st}$ order, we rule out the existence of a tricritical point in Cr for uniaxial pressures $p_{[110]} \le 550\,\mathrm{bar}$ despite the observation that the sample is apparently in a single $\bf Q_\pm$-state above $448\, {\rm bar}$ (Fig. \ref{NEUTRONfigPStep}). Reaching $550\, {\rm bar}$ is a significant improvement in comparison to previous work in which no broadening was observed up to only $160\, {\rm bar}$. \cite{BIBFawcettUniax} As a precondition for this conjecture, we succeeded to reduce the stress inhomogeneities strongly in our setup.

Comparing our results with a similar study on MnO by Bloch et al. \cite{bloch1975} we hesitate to rule out the appearance of a tricritical point in Cr under controlled uniaxial pressure exceeding $p_{[110]} > 550\,\mathrm{bar}$. Bloch et al. observed that a significant detwinning occured in MnO at $p = 1.2\, {\rm kbar}$ while clear evidence for tricritical behavior was only observed at a pressure $p = 5.5\, {\rm kbar}$. \cite{bloch1975}. Therefore, our uniaxial pressure device should be improved further to reach uniform pressures higher than $550\, {\rm bar}$. 

In contrast to Cr, where magnetostrictive effects are extremely small, i.e. $\triangle d/d \simeq 1\cdot10^{-5}$, \cite{combley1968} a strong spontaneous contraction of the lattice occurs in the ordered phase of antiferromagnetic MnO, i.e. the angle between the $[100]$ and the $[010]$ axis becomes $1.12\times 10^{-2}\, {\rm rad}$ at $4\, {\rm K}$. \cite{bloch1975} Therefore one may speculate that in the presence of the strongly enhanced magnetic fluctuations reported by Sternlieb et al. at the silent satellite positions in Cr close to $T_N$, \cite{sternlieb1995} and the extremely small magnetostrictive effects the phase space for magnetic fluctuations in Cr is not reduced when uniaxial pressure forces Cr to become single $\bf Q_\pm$. Hence, a stress-induced tricritical point in Cr may not exist and the phase transition in Cr remains weakly first-order.



\medskip

\begin{acknowledgements}

We gratefully acknowledge support by Reinhard Schwikowski, Andreas Mantwill and the team of FRM II for their technical support. We thank Georg Brandl and Enrico Faulhaber for their support with the computer infrastructure at MIRA. We greatfully acknowledge financial support through DFG Grant No. TRR80 ``From Electronic Correlations to Functionality''.


\end{acknowledgements}


\bibliography{paper_physrevb}



\end{document}